%% file: main.tex
\documentclass[conference,twocolumn]{IEEEtran}

\input{settings}

\bibliography{refs/formalmethods-FROZEN}

\begin{document}
\renewcommand{\figurename}{Figure}

\title{WIP: An Engaging Undergraduate Intro to Model Checking in Software Engineering Using \TLAplus}
\input{authors}
\maketitle

\TODO{grammar check by exporting plain text}
\TODO{attributions for microwave images}

\thispagestyle{plain}
\pagestyle{plain}




\newcommand{\structure}[1]{\textit{#1}: }

\begin{abstract}
\structure{Background} 
In this Innovative Practice Work in Progress, we present our initial efforts to integrate formal methods, with a focus on model-checking specifications written in Temporal Logic of Actions (\TLAplus), into computer science education, targeting undergraduate juniors/seniors and graduate students. Many safety-critical systems and services crucially depend on correct and reliable behavior. Formal methods can play a key role in ensuring correct and safe system behavior, yet remain underutilized in educational and industry contexts. \structure{Aims}
We aim to (1) qualitatively assess the state of formal methods in computer science programs, (2) construct level-appropriate examples that could be included midway into one's undergraduate studies, (3) demonstrate how to address successive ``failures'' through progressively stringent safety and liveness requirements, and (4) establish an ongoing framework for assessing interest and relevance among students. 
\structure{Methods}
We detail our pedagogical strategy for embedding \TLAplus into an intermediate course on formal methods at our institution. After starting with a refresher on mathematical logic, students specify the rules of simple puzzles in \TLAplus and use its included model checker (known as TLC) to find a solution. We gradually escalate to more complex, dynamic, event-driven systems, such as the control logic of a microwave oven, where students will study safety and liveness requirements. We subsequently discuss explicit concurrency, along with thread safety and deadlock avoidance, by modeling bounded counters and buffers. 
\structure{Results}
Our initial findings suggest that through careful curricular design and choice of examples and tools, it is possible to inspire and cultivate a new generation of software engineers proficient in formal methods. \structure{Conclusions} Our initial efforts suggest that 84\% of our students had a positive experience in our formal methods course. Our future plans include a longitudinal analysis within our own institution and proposals to partner with other institutions to explore the effectiveness of our open-source and open-access modules.
\end{abstract}




\section{Introduction}

In the evolving landscape of computer science education, methodologies that help to enhance the correctness and reliability of safety-critical systems are of vital importance~\cite{newcombe_how_2015,huisman_formal_2020}.
\GKT{Examples include those that George and Konstantin will discuss. I still remember Jeanette Wing coming to ANL to discuss Larch [theorem prover -> specification language tied to C++ https://www.cs.ucf.edu/~leavens/larchc++.html] Many examples are connected to hardware with embedded software and cyber-physical systems, e.g. drones, sensor-based systems [garage door?], failure-sensitive applications (distributed/cloud services), others?}
Recent advances in formal methods, particularly model checking of specifications written in \TLAplus~\cite{kuppe_tla_2019} have emerged as a powerful tool in bridging the gap between theoretical computer science and practical software engineering.
Although formal methods---rooted in automata theory, formal logic and, more specifically, Hoare logic~\cite{hoare_axiomatic_1969}---%
have a long history and continue to evolve, they have only modestly impacted CS education compared to many topics (e.g. machine learning, cybersecurity, Internet of Things, among others).
Although generative AI shows promise to reduce the error-prone nature of human coding, we remain convinced that humans---and non-humans---can employ formal methods to provide critical correctness and reliability checks on solutions, whether crafted by hand or generated by (statistical) AI and large-language ML models.\KL{OK to abbreviate here to save a line?}

\begin{figure}[t]
  \centering
  \includegraphics[width=0.6\linewidth]{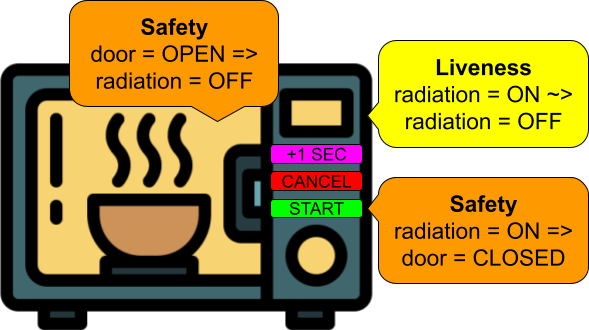}
  \caption{
  The microwave oven is an engaging and comprehensible, yet non-trivial, example of a safety-critical embedded system worthy of student attention in a Formal Methods course.
  This figure illustrates how we model two critical requirements:
    (1) To prevent radiation exposure, the appliance must \emph{not} run with the door open, and
    (2) To prevent overheating while radiating, it must eventually turn off (per the \TLAplus \emph{leads to} operator $\leadsto$).}
  \label{figure:COMP335Microwave}
  \vspace*{-3mm}
\end{figure}


In this paper, we describe our initial efforts to introduce students (undergraduate juniors/seniors and graduate students) to formal methods in software engineering with a focus on model checking using \TLAplus, backed by preliminary results from ongoing anonymous input from our own students.
We note that the selection of \TLAplus is intentional, as it is one of the most actively-maintained model checking systems (evidence provided in related work). 
We also incorporate other strategies and systems in our teaching: Students give presentations and demos on a broad landscape of tool- and language-based formal methods, including many of those shown in~\cref{tbl:RelatedCoursesAdHoc}.
\TODO{KL explain that this is a relevant embedded system with firmware}
By applying \TLAplus to practical, familiar, and compelling examples, such as the microwave oven shown in~\cref{figure:COMP335Microwave} and those listed in~\cref{tbl:COMP335SECourseExamples}, we aim to demystify formal methods and make them more relevant to students.

\GKT{More to be added here including a transition to the rest of the paper.}

\smallskip
\noindent
\ul{Our contributions include:}
\begin{itemize}
    \item Performing an initial analysis of academic programs offering formal methods courses, organized by course level, tools and techniques covered, and when last taught.
    \item Creating a set of curricular materials (i.e., exemplars) to motivate student interest in formal methods. We emphasize reproducible behavior using modern software engineering principles (i.e., version control and continuous integration) as our own materials and \TLAplus both continue to evolve.
    \item Demonstrating how to address successive ``failures'' in model checking through a set of progressively stringent safety and liveness requirements.
    \item Conducting an initial evaluation of the effectiveness of formal methods among our current students. This effort will inform ongoing study on the long-term effectiveness of our curricular approach.
\end{itemize}

\section{Background and Related Work}


The private computing industry and high-level government agencies have long recognized the value of formal methods for building safety-critical and other highly reliable systems~\cite{lecomte_applying_2020}.
A major example is the National Aeronautics and Space Administration (NASA), which ``develops formal methods technology for the development of mission-critical and safety-critical digital systems of interest to NASA~\cite{maddalon_nasa_2024}.''
Another safety-critical sector is ground transportation; a comprehensive survey of the use of formal methods in the railroad sector found that model checking is the most commonly adopted technique (47\%), followed by simulation (27\%) and theorem proving (19.5\%)~\cite{ferrari_formal_2022}. 
Specifically, \emph{model checking} arose in the early 1980s in response to the challenges of concurrent program verification~\cite{clarke_birth_2008} and is now well established and mature~\cite{clarke_formal_1996,grumberg_25_2008,clarke_handbook_2018,lecomte_applying_2020}.
Among various specification languages that support model checking, \TLAplus has emerged as a practical and effective option~\cite{lamport_teaching_2009,newcombe_how_2015,,ferrari_systematic_2022,kuppe_teaching_2023}.


The need to recruit qualified professionals in these areas became apparent several decades ago and is now receiving renewed attention~\cite{jaume_teaching_2014,mandrioli_heroism_2015,cerone_rooting_2020,huisman_formal_2020}.
Accordingly, academic educators have increasingly focused on formal methods teaching since the 1990s~\cite{clarke_formal_1996} and have been teaching model checking more broadly since the early 2000s~\cite{liu_proposal_2002,liu_teaching_2009,tahara_evolution_2009,tavolato_integrating_2012,schreiner_teaching_2018,aceto_introducing_2021,korner_increasing_2021,freiberger_model_2023}.
Several academic institutions have reported on their use of \TLAplus in their courses~\cite{mauran_teaching_2012,tavolato_integrating_2012}.

Nevertheless, multiple educators have recognized the challenges of motivating the need for formal methods topics and teaching them effectively in the face of some resistance~\cite{reed_motivating_2004,noble_more_2022}.
Key challenges include insufficient mathematical background from high school and university-level prerequisite courses, which may not focus sufficiently on discrete math and proof techniques; a lack of practical, engaging case studies that motivate the use of formal methods; insufficient documentation of some of the tools for students to gain confidence in using them; and learning styles of today's students (millennials and beyond), such as active, discovery-driven learning, an emphasis on solutions over theory, and a desire for immediate feedback~\cite{catano_engaging_2018,askarpour_teaching_2020}.

\KL{Which others to cite?
+reed_motivating_2004 - mental resistance, need small, engaging examples
zingaro_another_2008 - emphasize coupling of spec and impl in earlier courses
zamansky_mathematical_2016 - need to explain connection between logic and SE
korner_increasing_2021 - overwhelmed with transfer tasks
yatapanage_introducing_2021 - hate math, struggle with programming
+noble_more_2022 - perceived difficulty, excessive math content, practical irrelevance -> programming intensive approach based on Dafny
}

While some educators actually propose starting in the first year of the post-secondary curriculum~\cite{aceto_introducing_2021} or even earlier~\cite{gibson_formal_2008,yadav_introducing_2011}, we are concerned that these efforts will reach a relatively small number of students mostly at selective institutions. 
Therefore, we target juniors and seniors (at undergraduate level) and graduate students, and start with a three-week review of fundamentals studied earlier.




Our work builds on Askarpour and Bersani's study~\cite{askarpour_teaching_2020}, which not only reports on specific experiences at their institution, but also includes a general analysis of the Formal Methods Education Database (FMEDB)~\cite{ferreira_fme_2019} and a review of related work on the main challenges students face in formal methods courses.
We analyzed FMEDB for the most commonly used tools in 34 courses focused on model checking (out of 88 courses total) and found that 
7 use NuSMV/NuXMV, 
2 use Dafny, and 
2 use \TLAplus (one in industry and one in academia).
\KL{these are the results of our own web-search-based analysis of 17 courses}   
\GKT{Previously, we had this; however, I simplified to just focus on the results. We could still include this sentence but I wanted to make sure the numbers are right as they are not mentioned in Figure 2: The entire FME database includes 88 courses, 34 of which include model checking as a topic).}  
\KL{You lost the separation between FME and our own results - please do not edit without checking with me}
Because these entries are self-submitted and the database doesn't say when a course was last taught, we also conducted our own preliminary survey of relevant courses taught during the last five years, summarized in~\cref{tbl:RelatedCoursesAdHoc}.
As practicing software engineers, there are other reasons to use it: It is mature and well-maintained (by Leslie Lamport, a Turing Award winner) and has a strong community and ecosystem.
In sum, the overall related work strongly supports the decision to emphasize \TLAplus in our own course.

\begin{table}[htbp]
\caption{
Commonly used tools in FM courses, ranked by frequency of adoption as the primary tool, based on our preliminary survey of 32 recent courses; not shown are 15 courses without a primary tool or a less widely used one.
}
\label{tbl:RelatedCoursesAdHoc}
\centering
\footnotesize
\input{adhoc_courses_consolidated}
\end{table}

\section{Curricular Goals}

We target primarily undergraduates majoring in computer science or software engineering who have completed at least five foundational courses over three semesters: Introduction to Programming (CS1), Data Structures I (CS2), Introduction to Computer Systems (CS3), Discrete Structures, and a one-semester, hands-on introduction to the Linux command line.
In our semester-long, 15-week, three-credit course, we focus on the most relevant learning outcomes from the ACM/IEEE Computer Society Software Engineering 2014 Curriculum Guidelines~\cite{ardis_se_2015} in the order listed in~\cref{tbl:COMP335SEKnowledgeAreas}.


For many of our students, the most recent logic-based mathematics course is Discrete Structures, which most students take in their first year.
To aid in the transition to our course on Formal Methods, we reinforce the essential mathematical foundations. \TODO{(FND/MAA - elaborate)}
This material, however, is targeted toward what we need here, using unit testing as a vehicle to connect these foundations with the programming practice they have undergone in their CS1 and CS2 courses.

\GKT{Is there a week or so where you do any sort of refresher on the ideas of mathematical logic? I think we want to include something about this, based on the other related work that informs the idea. For many students, I'm guessing COMP 163 is the last time they saw any math related to this course.}
\KL{Yes, MAA.md.2 covers Hoare logic, or we could insert 3h of FND.mf.1/2: functions, relations, sets, predicate logic, and reduce testing from 12 to 9.}


\begin{table}
\caption{Formal Methods course: Approximate mapping from second-level knowledge areas to contact hours (15 weeks, 3 semester credits).}
\label{tbl:COMP335SEKnowledgeAreas}
\footnotesize
\centering
\begin{tblr}{|l|r|}
\hline
\textbf{SE 2014 Knowledge Area (Reference Code)} & \textbf{Hours} \\
\hline
\hline
Mathematical foundations (FND.mf) & 6 \\
\hline
Modeling foundations (MAA.md) & 3 \\
\hline
Testing (VAV.tst) & 9 \\
\hline
Types of models (MAA.tm) & 12 \\
\hline
Analysis fundamentals (MAA.af) & 6 \\
\hline
Requirements specification (REQ.rsd) & 3 \\
\hline
Developing secure software (SEC.dev) & 3 \\
\hline
Various professional practice topics (PRF) & 3 \\
\hline
\hline
\textbf{TOTAL} & 45 \\
\hline
\end{tblr}
\end{table}

We first build on students' understanding of correctness of a system under test w.r.t.\ a set of functional requirements and their ability to write automated, non-exhaustive unit tests that reflect these requirements. We also leverage students' understanding of discrete structures, especially predicate logic, to progress from example-based testing to universally quantified properties that more generally reflect the functional requirements, using libraries that automatically test these properties on a range of pseudo-randomly generated arguments. 

We then use \TLAplus to progressively introduce students to the formal specification and verification of dynamic and concurrent software systems. 
In \TLAplus, a model has a finite set of variables, and each state maps these variables to their current values.
\TLAplus allows us to model specific system behaviors that start in an initial state and undergo a sequence of states by performing transitions available in the next-state relation.
Beginning with intuitive examples, such as puzzles, we specify the rules in \TLAplus and use its included model checker (known as TLC) to falsify the absence of a solution, thereby finding a counterexample that constitutes a solution. 
We gradually escalate to more complex, dynamic systems, such as the control logic of a microwave oven, where students will study safety and liveness requirements; for greater engagement, we emphasize event-driven, interactive systems.
We subsequently study explicit concurrency, along with thread safety and deadlock avoidance, by modeling bounded counters and buffers. 
By proceeding gradually, we scaffold student understanding and application of formal methods (see~\cref{tbl:COMP335SECourseExamples}).

\begin{table}
\caption{Formal Methods course: Examples with targeted knowledge areas.
\GKT{Nicely done. Can we try to make our examples one-liners though. Space is at a real premium now. As an example, the Example is Palindrome checker, the Knowledge Area is Unit Testing. Keep it concise. Whether stateful or stateless, I think these details will be lost on most readers. The caption could have words stateful and stateless but is more of a detail, IMHO.}
}
\label{tbl:COMP335SECourseExamples}
\footnotesize
\centering
\begin{tblr}{|l|l|l|}
\hline
\textbf{Example} & \textbf{Tool} & \textbf{Knowledge Areas} \\
\hline
\hline
{Unit testing: palindrome checker}& JUnit & Testing \\\hline
{Property-based testing: palindrome} & jqwik & Testing \\\hline
{Stateful testing: circular buffer} & jqwik & Testing \\\hline
Microwave oven (see~\cref{sec:CourseActivities}) & \TLAplus & Modeling, Requirements \\\hline
Elevator control logic & \TLAplus & Modeling, Requirements \\\hline
Shared counter, explicit threads & \TLAplus & Modeling, Concurrency \\\hline
Bounded buffer, explicit threads & \TLAplus & Modeling, Concurrency \\\hline
\end{tblr}
\end{table}


\section{Course Activities}
\label{sec:CourseActivities}

Here we present a series of activities for students based on a simple microwave oven as a running example.

\subsection{Running Example: A Simple Microwave Oven}


We represent the state of our microwave as two booleans, \textit{door} (open or closed) and \textit{radiation} (on or off), along with a natural number, \textit{timeRemaining}. 
The system also supports various actions, mostly corresponding to user controls:
\begin{itemize}
\item The user can open or close the door.
\item The user can increment the remaining time.
\item The user can start or cancel the heating process.
\item An internal timer decrements the remaining time to zero.
\end{itemize}

As seen in~\cref{fig:MicrowaveInit}, radiation is initially off with no time remaining, while the door may be open or closed.
The model's next-state relation \textit{Next} is defined as the logical disjunction (choice) of the user-triggered actions \textit{OpenDoor}, \textit{CloseDoor}, \textit{IncTime}, \textit{Start}, \textit{Cancel}, as well as the internal \textit{Tick} action. 

\begin{figure}[htbp]
\footnotesize
\input{snippets/snip-tla-microwave-init}
\smallskip
\caption{
Valid initial states and top-level specification for the microwave oven: 
Initially, the microwave is not radiating, there is no time remaining, and the door is either open or closed. 
Subsequently, the model can repeatedly perform any available action, as indicated by the temporal operator ``always'' ($\Box$).
}
\label{fig:MicrowaveInit}
\end{figure}

\textbf{Activity 1} Starting with a skeletal model where each action manually controls the corresponding variable, develop appropriate state-dependent behavior, such as the timer counting down and automatically shutting off radiation when reaching zero. 
\cref{fig:MicrowaveTick} defines the resulting \textit{Tick} action, and~\cref{fig:MicrowaveScenarioNormal} shows a step in a normal behavior.

\begin{figure}[htbp]
\footnotesize
\input{snippets/snip-tla-microwave-tick}
\smallskip
\caption{
\textit{Tick} action for decrementing the timer, defined as a conjunction of preconditions, effects, and postconditions, where $v'$ refers to a variable's value in the next state following the action.
According to the preconditions, this action is enabled only when the oven is radiating and there is nonzero time remaining. 
Its effect is to decrement the remaining time; 
when the remaining time reaches zero, its additional effect is to shut off radiation.
}
\label{fig:MicrowaveTick}
\end{figure}

\begin{figure}[htbp]
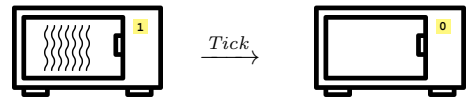

\vspace*{-5mm}
\centering
\begin{tblr}{colspec={Q[l,t]Q[c,b]Q[r,t]}}
\MicrowaveState{c}{r}{1}
&
{$\xrightarrow{Tick}\ $\\[\transitionarrowvspace]}
&
\MicrowaveState{c}{o}{0}
\end{tblr}
\vspace*{-3mm}
\caption{Scenario for normal operation: 
The microwave is initially radiating with the door closed and one second of time remaining. 
When the internal timer ticks, the remaining time goes to zero, and the oven stops radiating.
}
\label{fig:MicrowaveScenarioNormal}
\end{figure}

\subsection{Invariants and Safety}

To study system safety, we will initially allow the door and radiation to operate independently.
The TLC model checker finds nothing wrong with this model because we have not yet defined any invariants it can attempt to falsify.

\textbf{Exercise 2a} Define an invariant \textit{DoorSafety} to guarantee that the microwave will never be radiating with the door open.
What happens after you add this invariant?

\begin{tlatex}
\footnotesize
\@pvspace{4.0pt}%
\@x{\@s{8.2} DoorSafety \.{\defeq} door \.{=} OPEN \.{\implies} radiation \.{=} OFF}
\@pvspace{4.0pt}%
\end{tlatex}

\begin{figure}[htbp]
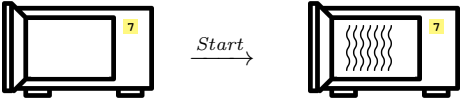

\vspace*{-5mm}
\centering
\begin{tblr}{colspec={Q[l,t]Q[c,b]Q[r,t]}}
\MicrowaveState{o}{o}{7}
&
{$\xrightarrow{Start}\ $\\[\transitionarrowvspace]}
&
\MicrowaveState{o}{r}{7}
\end{tblr}
\vspace*{-3mm}
\caption{Scenario leading to an unsafe condition: 
The microwave is initially stopped with the door open and several seconds of time remaining.
When the user presses the start button, the microwave emits harmful radiation.
}
\label{fig:MicrowaveScenarioStart}
\end{figure}

\noindent
Once we add this invariant to our model, TLC considers \emph{unsafe} any state in which the invariant is false and proves that there is at least one behavior leading to such a state (see~\cref{fig:MicrowaveScenarioStart}).

\textbf{Exercise 2b} Refine the microwave model to satisfy the \textit{DoorSafety} invariant, making the smallest possible changes.

This requires introducing more state dependency within the actions constituting the next-state relation.
To prevent the scenario from~\cref{fig:MicrowaveScenarioStart}, we can require the door to be closed by adding a precondition to the \textit{Start} action:

\begin{tlatex}
\footnotesize%
\@pvspace{4.0pt}%
\@x{\@s{8.2} \.{\land} door \.{=} CLOSED}%
\@pvspace{4.0pt}%
\end{tlatex}

\begin{figure}[htbp]
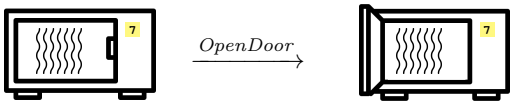

\vspace*{-5mm}
\centering
\begin{tblr}{colspec={Q[l,t]Q[c,b]Q[r,t]}}
\MicrowaveState{c}{r}{7}
&
{$\xrightarrow{OpenDoor}\ $\\[\transitionarrowvspace]}
&
\MicrowaveState{o}{r}{7}
\end{tblr}
\vspace*{-3mm}
\caption{Scenario leading to an unsafe condition: 
The microwave is initially radiating with the door closed and several seconds of time remaining.
When the user opens the door, harmful radiation comes out.
}
\label{fig:MicrowaveScenarioOpen}
\end{figure}

\noindent
But TLC quickly finds another unsafe behavior, as shown in~\cref{fig:MicrowaveScenarioOpen}.
To prevent this scenario, we will add to the \textit{OpenDoor} action the effect of immediately shutting off radiation.

\begin{tlatex}
\footnotesize
\@pvspace{4.0pt}%
\@x{\@s{8.2} \.{\land} radiation \.{'} \.{=} OFF}%
\@pvspace{4.0pt}%
\end{tlatex}

\subsection{Temporal Properties and Liveness}

Our microwave model is now safe in terms of preventing any behaviors that allow radiation to occur when the door is open.
\emph{But is this notion of safety enough?} How do we know it won't radiate indefinitely, overheat, and catch on fire?

\TLAplus includes temporal logic operators for predicates on a sequence of steps, e.g., some false condition eventually becoming true.
Temporal logic predicates are often used to express liveness requirements, such as a radiating microwave oven eventually reaching zero remaining time and shutting off.

\textbf{Exercise 3a} Define a temporal property \textit{HeatLiveness} requiring a radiating microwave to eventually turn off.
What happens after you add this property?

\begin{tlatex}
\footnotesize
\@pvspace{4.0pt}%
\@x{\@s{8.2} HeatLiveness \.{\defeq} radiation \.{=} ON \.{\leadsto} radiation \.{=} OFF}
\@pvspace{4.0pt}%
\end{tlatex}

\noindent
The ``leads to'' operator in $p \leadsto q$ indicates that if $p$ is currently true, then $q$ must become true in a future step.

\begin{figure}[htbp]
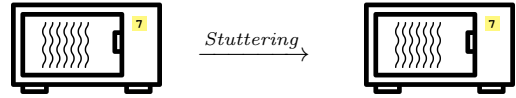

\vspace*{-5mm}
\centering
\begin{tblr}{colspec={Q[l,t]Q[c,b]Q[r,t]}}
\MicrowaveState{c}{r}{7}
&
{$\xrightarrow{Stuttering}\ $\\[\transitionarrowvspace]}
&
\MicrowaveState{c}{r}{7}
\end{tblr}
\vspace*{-3mm}
\caption{Scenario showing lack of liveness in the form of stuttering: The microwave is radiating but not receiving ticks required to reach zero.}
\label{fig:MicrowaveScenarioStuttering}
\end{figure}

Once we add this liveness property to our model, TLC will indicate that our current model does allow the scenario from~\cref{fig:MicrowaveScenarioStuttering}.
Even though this behavior involves only valid states, it is undesirable because it allows radiation to continue indefinitely and the food inside the microwave to catch on fire.

In general, \TLAplus does not require the model to choose an action even if one is available. 
Instead, the model is allowed to perform a transition where it stays in its current state; this is called a \emph{stuttering} step.
This can model scenarios where a system loses power or, as in~\cref{fig:MicrowaveScenarioStuttering}, fails to make progress. 

\smallskip
\textbf{Exercise 3b} Refine the model to satisfy the \textit{HeatLiveness} temporal property, making the smallest possible changes.

To break the observed stutter-invariance, we need to use a mechanism called \emph{(weak) fairness}.
We can enable this for the \textit{Tick} action by adding $\land {\WF}_{vars}({Tick})$ to our top-level specification.
The resulting model satisfies our previously established safety and liveness requirements.

\section{Preliminary Evaluation} 


We offered the course for the first time in the Fall 2022 semester to 15 students, and again in the Spring 2024 semester to 22 students. 
Our new course-specific 15-question survey, given as a pretest-posttest in Spring 2024, indicates a marked improvement in student receptiveness to formal methods, with enhanced ability to conceptualize and apply these techniques in real-world scenarios; 
the cohort average of equally weighted composite scores increased from 3.0 to 4.1 ($n = 19$), with 84\% reporting a positive experience.
Notably, the use of \TLAplus has not only facilitated a deeper understanding of software correctness but also stimulated student interest in the subject matter.
In addition, preliminary results from standardized teacher-course evaluations are highly encouraging, with average scores significantly above departmental averages and relatively high consensus among respondents. 
\KL{For later? In addition, standardized teacher-course evaluations for the Spring 2024 offering of this course were very positive, with average scores significantly above departmental averages and relatively high consensus among respondents; while acknowledging the complexity of the material, students' written comments were also overwhelmingly positive.}




\section{Conclusion}

In summary, this work in progress demonstrates the successful integration of model checking using \TLAplus into our intermediate undergraduate computer science curriculum.
With just three semesters (five courses) of prerequisites, students can learn about formal methods and model checking as early as their second semester of sophomore year and most certainly by the first semester of junior year, grounded in the ACM/IEEE Computing Curricula, which strongly suggest what students should learn and when they should learn it, including all appropriate foundational preparation.
With our open source and open access materials (code and lecture) notes continuously updated on GitHub and rebuilt and tested using continuous integration, anyone at any institution can use and contribute to our work.
Our initial findings strongly suggest that this approach, including the use of effective, well-documented tools such as \TLAplus, is engaging to students.

Our future plans include reaching out to other universities, starting with our own region, which has dozens of universities, and potentially offering a workshop aimed at workforce development for those who may have missed the opportunity to learn this material while studying computer science or a closely-related discipline.


\section*{Acknowledgments}
{
This material is based upon work supported by the National Science Foundation under Grant No.\ DGE-1919004. Any opinions, findings, and conclusions or recommendations expressed in this material are those of the author(s) and do not necessarily reflect the views of the National Science Foundation.
The authors thank Ryan Hasler for helpful suggestions for improving the included \TLAplus examples.
}


\printbibliography

\end{document}

%% file: settings.tex
\usepackage[style=ieee,doi=false,isbn=false,url=false,eprint=false]{biblatex}

\usepackage{comment}
\usepackage{algorithmic}
\usepackage{graphicx}
\usepackage{textcomp}
\usepackage{xcolor}
\usepackage{tabularray}
\usepackage{booktabs} 
\usepackage{xspace} 
\usepackage[normalem]{ulem}
\usepackage{makecell}
\usepackage{hyperref}
\usepackage{relsize} 

\usepackage{svg}
\svgpath{figures}
\usepackage{outlines} 
\usepackage[most]{tcolorbox}

\usepackage{balance} 
\usepackage{soul}
\usepackage{cleveref}
\usepackage{booktabs}
\usepackage{lipsum}
\usepackage{adjustbox}
\usepackage{tabularx}
\usepackage{paralist}
\usepackage[frozencache=true,cachedir=minted-cache]{minted}

\usepackage{tlatex}

\usepackage{tikz}

\usepackage{framed}
\definecolor{shadecolor}{gray}{0.97}

\crefformat{section}{\S#2#1#3}
\crefname{figure}{Figure}{Figures}
\crefname{appendix}{Appendix}{Appendices}
\crefname{table}{Table}{Tables}
\crefname{algorithm}{Algorithm}{Algorithms}
\crefname{listing}{Listing}{Listings}
\crefname{theorem}{Theorem}{Theorems}
\crefname{thm}{Theorem}{Theorems}
\crefname{lemma}{Lemma}{Lemmata}
\crefname{equation}{Eqt.}{Eqts.}
\crefformat{Grammar}{Grammar #1}

\ifCLASSOPTIONcompsoc
  \usepackage[caption=false,font=normalsize,labelfont=sf,textfont=sf]{subfig}
\else
  \usepackage[caption=false,font=footnotesize]{subfig}
\fi

\newif\ifDEBUG
\DEBUGfalse
\newif\ifANONYMOUS
\ANONYMOUStrue

\setlipsum{%
  par-before = \begingroup\color{gray},
  par-after = \endgroup
}

\ifDEBUG
    \newcommand{\GKT}[1]{\textcolor{brown}{[GKT: #1]}}
    \newcommand{\KL}[1]{\textcolor{blue}{[KL: #1]}}
    \newcommand{\TODO}[1]{\hl{#1}}
\else
    \newcommand{\GKT}[1]{}
    \newcommand{\KL}[1]{}
    \newcommand{\TODO}[1]{}
\fi

\usepackage{etoolbox}
\makeatletter
\patchcmd{\ttlh@hang}{\parindent\z@}{\parindent\z@\leavevmode}{}{}
\patchcmd{\ttlh@hang}{\noindent}{}{}{}

\patchcmd{\@makecaption}
  {\scshape}
  {}
  {}
  {}
\makeatother

\newcolumntype{R}[2]{%
    >{\adjustbox{angle=#1,lap=\width-(#2)}\bgroup}%
    l%
    <{\egroup}%
}

\IEEEoverridecommandlockouts

\definecolor{cellbackground}{HTML}{F7F7F7}

\newminted{java}{bgcolor=cellbackground}
\newminted{scala}{bgcolor=cellbackground}
\newminted{c}{bgcolor=cellbackground}
\newminted{cpp}{bgcolor=cellbackground}
\newminted{bash}{bgcolor=cellbackground}
\newminted{cmake}{bgcolor=cellbackground}
\newminted{text}{bgcolor=cellbackground}

\setminted[scala]{bgcolor=cellbackground}
\setminted[scala]{fontsize=\footnotesize}
\setminted[scala]{autogobble} 
\setminted[java]{bgcolor=cellbackground}
\setminted[java]{fontsize=\footnotesize}
\setminted[java]{autogobble} 
\setminted[c]{bgcolor=cellbackground}
\setminted[c]{fontsize=\footnotesize}
\setminted[c]{autogobble} 
\setminted[cpp]{bgcolor=cellbackground}
\setminted[cpp]{fontsize=\footnotesize}
\setminted[cpp]{autogobble}
\setminted[bash]{bgcolor=cellbackground}
\setminted[bash]{fontsize=\footnotesize}
\setminted[bash]{autogobble}
\setminted[cmake]{bgcolor=cellbackground}
\setminted[cmake]{fontsize=\footnotesize}
\setminted[cmake]{autogobble}
\setminted[text]{bgcolor=cellbackground}
\setminted[text]{fontsize=\footnotesize}
\setminted[text]{autogobble}

\newmintinline{c}{showspaces}

\newmintinline[cvar]{c}{showspaces}
\newmintinline[cconst]{c}{showspaces}
\newmintinline[cfunction]{c}{showspaces}
\newmintinline[ctype]{c}{showspaces}
\newmintinline[cexpr]{c}{showspaces}

\newcommand{\TLAplus}{TLA\textsuperscript{+}\xspace}

\newcommand\MicrowaveState[3]{
\scalebox{0.4}{
\begin{tikzpicture}
  \ifx o#1
    \node[anchor=south west,inner sep=0] (image) at (0,0) {\includegraphics[width=5cm]{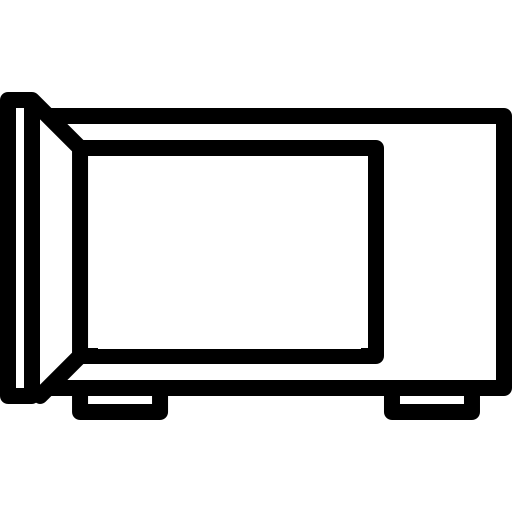}};
    \begin{scope}[x={(image.south east)},y={(image.north west)}]
        \node[fill=yellow, opacity=0.5, text opacity=1] at (0.85,0.65) {\large\textbf{\texttt{#3}}\relax};
    \end{scope}
    \ifx r#2
      \node[anchor=south west,inner sep=0] (image) at (1.25,1.8) {\includegraphics[width=1.5cm]{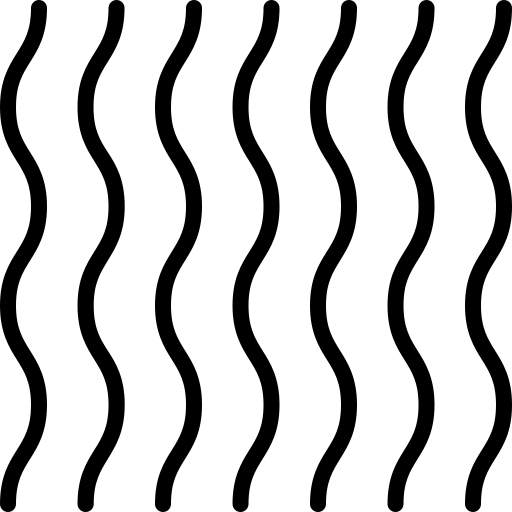}};
    \fi
  \else
    \node[anchor=south west,inner sep=0] (image) at (0,0) {\includegraphics[width=5cm]{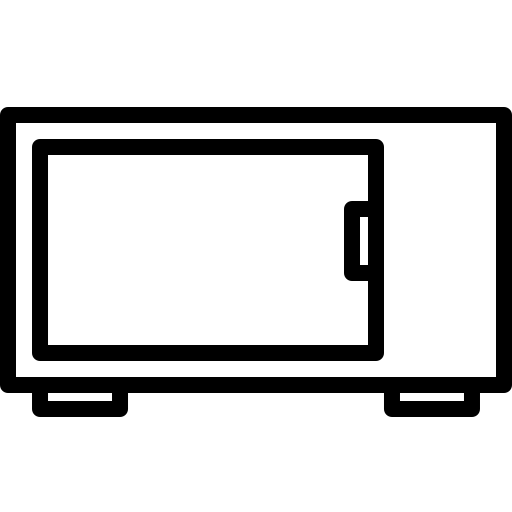}};
    \begin{scope}[x={(image.south east)},y={(image.north west)}]
        \node[fill=yellow, opacity=0.5, text opacity=1] at (0.85,0.65) {\large\textbf{\texttt{#3}}\relax};
    \end{scope}
    \ifx r#2
      \node[anchor=south west,inner sep=0] (image) at (1.05,1.8) {\includegraphics[width=1.5cm]{figures/waves.png}};
    \fi
  \fi
\end{tikzpicture}
}}

\newcommand\transitionarrowvspace{8mm}

\crefformat{section}{\S#2#1#3}
\crefname{figure}{Figure}{Figures}
\crefname{appendix}{Appendix}{Appendices}
\crefname{table}{Table}{Tables}
\crefname{algorithm}{Algorithm}{Algorithms}
\crefname{listing}{Listing}{Listings}
\crefname{theorem}{Theorem}{Theorems}
\crefname{thm}{Theorem}{Theorems}
\crefname{lemma}{Lemma}{Lemmata}
\crefname{equation}{Eqt.}{Eqts.}
\crefformat{Grammar}{Grammar #1}


%% file: authors.tex

\author{
  \IEEEauthorblockN{
    Konstantin Läufer\IEEEauthorrefmark{1},
    Gunda Mertin\IEEEauthorrefmark{2},
    and
    George K.\ Thiruvathukal\IEEEauthorrefmark{1}
  }
  \IEEEauthorblockA{
    \IEEEauthorrefmark{1}Software and Systems Laboratory, 
      Department of Computer Science,
      Loyola University Chicago\\
      \{laufer,gkt\}@cs.luc.edu
    \\
    \IEEEauthorrefmark{2}Institute for Software Engineering and Programming Languages,
      University of Lübeck\\
      gunda.mertin@student.uni-luebeck.de
  }
}

%% file: adhoc_courses_consolidated.tex
\begin{tblr}{width=\linewidth,colspec={|l|c|c|}}
\hline
Tools&Number of Courses&Level\\\hline
\hline
TLA+&5&Grad\\\hline
Alloy&4&Both\\\hline
SPIN&4&Both\\\hline
Coq&2&Both\\\hline
Z3&2&Grad\\\hline
\end{tblr}

%% file: snippets/snip-tla-microwave-init.tex
\begin{tlatex}
\@x{ Init \.{\defeq} \.{\land} door \.{\in} \{ OPEN ,\, CLOSED \}}%
\@x{\@s{36.2} \.{\land} radiation \.{=} OFF}%
\@x{\@s{36.2} \.{\land} timeRemaining \.{=} 0}%
\@pvspace{8.0pt}%
\@x{ Spec \.{\defeq} Init \.{\land} {\Box} [ Next ]_{ vars}}%
\end{tlatex}

%% file: snippets/snip-tla-microwave-tick.tex
\begin{tlatex}
\@x{ Tick \.{\defeq} \.{\land} radiation \.{=} ON}%
\@x{\@s{40.2} \.{\land} timeRemaining \.{'} \.{=} timeRemaining \.{-} 1}%
\@x{\@s{40.2} \.{\land} timeRemaining \.{'} \.{\geq} 0}%
\@x{\@s{40.2} \.{\land} {\IF} timeRemaining \.{'} \.{=} 0}%
\@x{\@s{48.2} \.{\THEN} radiation \.{'} \.{=} OFF}%
\@x{\@s{48.2} \.{\ELSE} {\UNCHANGED} {\langle} radiation {\rangle}}%
\@x{\@s{40.2} \.{\land} {\UNCHANGED} {\langle} door {\rangle}}%
\end{tlatex}